\documentclass[12pt]{iopart}
\usepackage{epsfig}
\usepackage{iopams}

\newcommand{\be}{\begin{equation}}
\newcommand{\ee}{\end{equation}}

\newcommand{\ba}{\begin{array}}
\newcommand{\ea}{\end{array}}
\newcommand{\bea}{\begin{eqnarray}}
\newcommand{\eea}{\end{eqnarray}}

\newcommand{\g}{\mathfrak{g}}
\newcommand{\glr}{sl(2, \mathbb{R})}

\newcommand{\nn}{\nonumber \\}
\newcommand{\prw}{pr \vec{v}_Q }

\begin{document}

\title{Soliton surfaces associated with symmetries of ODEs written in Lax representation}
\author{ A M Grundland$^1$ $^2$ and S Post$^1$}

\address{$^1$ Centre de Recherches Math\'ematiques. Universit\'e de Montr\'eal. Montr\'eal CP6128 (QC) H3C 3J7, Canada}
\address{$^2$ Department of Mathematics and Computer Sciences, Universit\'e du Quebec, Trois-Rivi\'eres. CP500 (QC)G9A 5H7, Canada}
\ead{grundlan@crm.umontreal.ca, post@crm.umontreal.ca}

\begin{abstract} The main aim of this paper is to discuss recent results on the adaptation of  the Fokas-Gel'fand procedure for constructing soliton surfaces in Lie algebras, which was originally derived for PDEs [Grundland, Post 2011], to the case of integrable ODEs admitting Lax representations. We give explicit forms of the $\g$-valued immersion functions based on conformal symmetries involving the spectral parameter, a gauge transformation of the wave function and generalized symmetries of the linear spectral problem. The procedure is applied to a symmetry reduction of the static $\phi^4$-field equations leading to the Jacobian elliptic equation. As examples, we obtain diverse types of surfaces for different choices of Jacobian elliptic functions for a range of values of parameters. 
\end{abstract}
\maketitle

\section{Introduction}\label{intro}
The methodological approach assumed in this work is based on the Fokas-Gel'fand formula for immersion of 2D-soliton surfaces associated with integrable models. This topic has been extensively developed by several authors (see e.g. \cite{Cies1997, FG, FGFL, Sym, Tafel}) and also by the authors \cite{GrundPost2011a} for integrable partial differential equations (PDEs). The results obtained proved to be so fruitful from the point of view of constructing 2D-surfaces immersed in Lie algebras that it seemed  worthwhile  to adapt this method and check its effectiveness for the case when an integrable ODE can be written in the Lax representation. The motivation for such an analysis is that it allows for the study of PDE surfaces using ODE surfaces as approximations. The construction of such soliton surfaces using the Fokas-Gel'fand approach is presented in \cite{GrundPost2011b} and its application to the symmetry reduction of the static $\phi^4$-field equations is the goal of this work.

\section{Soliton surfaces associated with integrable ODEs}\label{prelim}
Let us consider an ODE ($x$ stands for the independent variable and $u$ for the unknown function)
\be \label{delta} \Delta[u]\equiv \Delta(x,u,  u_x, u_{xx}, \ldots)=0\ee which   admits a Lax pair with potential matrices  $L(\lambda,[u]), \ M(\lambda,[u])$, taking values in a Lie algebra $\g,$ which satisfy
\be \label{cc} D_x M+[M,L]=0, \mbox{ whenever } \Delta[u]=0,\ee
where  $L(\lambda,[u]), \ M(\lambda,[u])$ are rational functions of a spectral parameter $\lambda$  taking values in either $\mathbb{C}$ or $\mathbb{R}.$

 In what follows, we make use of jet space and the prolongation structure of vector fields as presented in the book by P. J. Olver \cite{Olver}. For derivatives of $u$ we use the standard notation $u_x$ and $u_J$ for the first and $J$th derivatives of $u$ with respect to $x$, respectively. Functions depending on the independent variable $x,$ dependent variable $u$ and its derivatives are denoted
\[ f[u]=f(x,u, u_x, u_{xx}, \ldots)\]
and the total derivative in the direction of $x$ takes the form
\be D_x=\partial_x+u_x\frac{\partial }{\partial u} +u_{xx} \frac{\partial }{\partial u_x}+\ldots .\ee

This Lax pair equation \eref{cc} can be regarded as the compatibility conditions of a linear spectral problem (LSP) for the wave function $\Phi(\lambda, y, [u])$ taking values in the Lie group $G$ 
\be\label{lsp} \fl D_x\Phi(\lambda, y, [u]) = L(\lambda, [u])\Phi(\lambda, y, [u]), \qquad D_y \Phi(\lambda, y, [u]) =M(\lambda, [u])\Phi(\lambda, y, [u]).\ee 
Note that the wave function $\Phi$ depends on an auxiliary variable $y$ in the LSP \eref{lsp} and so the compatibility condition for \eref{lsp} coincides with \eref{cc} since $u_y=0.$
The total derivative in the direction $y$ is given  by 
\[ D_y=\frac{\partial}{\partial y}\]
and consequently, we obtain 
\be \label{Dy0} D_yL=D_yM=0.\ee

It was shown in \cite{FG}, that for any $\g$-valued functions $A(\lambda, y, [u])$ and $B(\lambda, y, [u])$ which satisfy 
\be \label{AB} D_y A-D_x B+[A,M]+[L, B]=0\ee
there exists a $\g$-valued function $F$ with tangent vectors given by 
\be \label{tanF} D_xF= \Phi^{-1} A \Phi, \qquad D_yF=\Phi^{-1} B \Phi.\ee
Whenever the matrices $A$ and $B$ are linearly independent, $F$ is an immersion function for a 2D surface in the Lie algebra $\g.$ It was proved in \cite{FG, FGFL, GrundPost2011a} that there exist three linearly independent terms satisfying \eref{AB} given by 
\bea \label{A} A&=a(\lambda) \frac{\partial \Phi}{\partial \lambda} L+D_x S +[S,L]+\prw L \in \g,\\
\label{B} B&=a(\lambda) \frac{\partial \Phi}{\partial \lambda}  M+D_y S +[S,M]+\prw M \in\g,
\eea
where $a(\lambda)$ is an arbitary scalar function of $\lambda$, $S=S(\lambda, y, [u])$ is an arbitrary $\g$-valued function, $\vec{v}_Q$ is a generalized symmetry of \eref{delta} and its prolongation is given by 
\be \vec{v}_Q=Q[u]\frac{\partial }{\partial u}, \qquad pr\vec{v}_Q=\vec{v}_Q+D_J(Q)\frac{\partial }{\partial u_J}.\ee
Here $\vec{v}_Q$ is assumed to be a generalized symmetry of a nondegenerate ODE \eref{delta}, i.e. 
\be pr\vec{v}_Q(\Delta[u])=0,\qquad \mbox{ whenever } \Delta[u]=0.\ee
Further, it was shown in \cite{GrundPost2011a} that the $\g$-valued function $F$ can be integrated (up to an additive $\g$-valued constant as)
\be\label{Ffg} F=a(\lambda)\Phi^{-1}\frac{\partial \Phi}{\partial \lambda} +\Phi^{-1} S\Phi +\Phi^{-1} \prw \Phi\in \g, \ee
as long as  $\vec{v}_Q$ is a generalized symmetry of the LSP \eref{lsp} and the ODE in Lax representation \eref{delta} in the sense that the following equations hold
\bea pr\vec{v}_Q\left(D_xM+[M,L]\right)=0, \qquad \mbox{ whenever } D_xM+[M,L]=0,\\
        pr\vec{v}_Q\left(D_x\Phi-L\Phi\right)=0, \qquad \mbox{ whenever } D_x\Phi-L\Phi=0,\\
        pr\vec{v}_Q\left(D_y\Phi-M\Phi\right)=0, \qquad \mbox{ whenever } D_y\Phi-M\Phi=0.\eea
The three terms in \eref{AB} correspond to a conformal transformation in the spectral parameter (known as the Sym-Tafel formula for immersion \cite{Sym, Tafel}), a gauge symmetry of the LSP \eref{lsp} and generalized symmetries of the ODE \eref{delta} and the LSP \eref{lsp}. The integrated form \eref{Ffg} define a mapping $F: \mathbb{R}^2\rightarrow \g$ which is called the Fokas-Gel'fand formula for immersion in a Lie algebra $\g$. In what follows, we will refer to it as such. 

As first indicated in \cite{FG} and extended in \cite{GrundPost2012}, there are many more choices of $A$ and $B$ which satisfy \eref{AB}. In fact, as proven in \cite{GrundPost2012}, any $\g$-valued function on jet space can be transformed into a symmetry of the Lax equation \eref{cc} when considered as an autonomous system of PDEs in the variables $L$ and $M$. For example, this system is invariant under the following point transformations: translation in the $x$ direction and conformal transformations in the spectral parameter $\lambda$ corresponding to terms associated with scalar constants $\alpha_1$ and $\alpha_2$ in \eref{Aext} and \eref{Bext}. The symmetry of the equation with respect to an arbitrary  gauge $S(\lambda, y, [u])\in \g$ for the wave functions $\Phi$ of the LSP \eref{lsp} is given by the term associated with $\alpha_3.$ The expressions corresponding to $\alpha_4$ and $\alpha_5$ are related to the invariance of \eref{delta} under dilations (i.e. $x\rightarrow e^\mu x, L\rightarrow e^{-\mu}L$ and  $y\rightarrow e^\mu y, M\rightarrow e^{-\mu}M$, $\mu \in \mathbb{R}$).
This leads to the following extension of the matrices $A,B\in \g$ given by \eref{A} and \eref{B} to 
\bea \label{Aext} \fl A&=\alpha_1D_x L+\alpha_2\frac{\partial L}{\partial \lambda}+\alpha_3(D_x S +[S,L]) +\alpha_4 D_x(xL)+\alpha_6\prw L ,\\
\label{Bext}\fl  B&=\alpha_1 D_x M +\alpha_2 \frac{\partial M}{\partial \lambda} +\alpha_3(D_y S +[S,M])+\alpha_4 xD_xM +\alpha_5D_y(yM)+\alpha_6\prw M .
\eea 

The case where $\alpha_1=\alpha_3=\alpha_4=\alpha_5=\alpha_6=0,$ $\alpha_2=a(\lambda)$ corresponds to the Sym-Tafel formula for immersion \cite{Sym, Tafel}, which is given by 
\be F^{ST}=a(\lambda)\Phi^{-1}\frac{\partial}{\partial \lambda} \Phi \in \g\ee
with tangent vectors 
\be \label{DFST} D_xF^{ST}=a(\lambda)\Phi^{-1} \frac{\partial \Phi}{\partial \lambda} L\Phi, \qquad 
       D_yF^{ST}=a(\lambda)\Phi^{-1} \frac{\partial \Phi}{\partial \lambda} M\Phi.\ee
If the tangent vectors \eref{DFST} are linearly independent, then the function $F^{ST}$ is an immersion of a 2D-surface in the Lie algebra $\g$. 

The case where $\alpha_1=\alpha_2=\alpha_4=\alpha_5=\alpha_6=0$ was studied in \cite{Cies1997, FGFL, GrundPost2011a} and the corresponding surfaces can be integrated explicitly as 
\be F^{S}=\Phi^{-1} S(\lambda, y, [u])\Phi\in \g\ee
with tangent vectors 
\be\label{DFS} D_xF^{S}=\Phi^{-1}\left(D_x S+[S,L]\right)\Phi, \qquad 
       D_yF^{S}=\Phi^{-1} \left(D_y S+[S,M]\right)\Phi.\ee
For $F^{S}$ to be an immersion, we require the linear independence of the tangent vector fields. 

Here, we show only the proofs for the case involving translation in the $x$ direction and the terms involving dilations. In the first case, (when  $\alpha_2=\alpha_3=\alpha_4=\alpha_5=\alpha_6=0$), the tangent vectors are given by 
\bea\label{DxF1} D_x F=\Phi^{-1} (D_xL)\Phi, \qquad A=D_xL,\\
       \label{DyF1} D_y F=\Phi^{-1} (D_xM)\Phi, \qquad B=D_xM.\eea
The matrices $A$ and $B$ satisfy the condition \eref{AB} 
\bea D_y(D_xL)-D_x(D_xM)+[D_xL,M]+[L,D_xM]\nn
=-D_x\left(D_xM+[M,L]\right)=0,\nonumber\eea
whenever the Lax equation \eref{cc} holds. Thus, this proves that there exists a $\g$-valued function $F$ with tangent vectors given by \eref{DxF1} and \eref{DyF1}. Furthermore, the immersion function can be integrated and is given by
\be\label{F1} F=\Phi^{-1} D_x\Phi=\Phi^{-1} L\Phi\ee
whenever the wave function $\Phi$ is a solution of the LSP \eref{lsp}.
It is straightforward to check that
\bea D_xF&=&-\Phi^{-1} (D_x\Phi) \Phi^{-1}L\Phi+\Phi^{-1}D_x(L\Phi)\nn
		&=&-\Phi^{-1} L^2\Phi+\Phi^{-1} D_xL \Phi +\Phi^{-1} L^2\Phi\nn
		&=&\Phi^{-1} (D_xL)\Phi\nonumber,\eea
		and 
\bea D_yF&=&-\Phi^{-1} (D_y\Phi) \Phi^{-1}L\Phi+\Phi^{-1}D_y(L\Phi)\nn
		&=&-\Phi^{-1}[L,M]\Phi\nn
		&=&\Phi^{-1} (D_xM)\Phi\nonumber,\eea
whenever \eref{cc} holds. Here we have used $D_yL=0.$ Hence, we have proved that the immersion function $F$ can be integrated as \eref{F1}. 

In the case of dilation symmetry (where $\alpha_1=\alpha_2=\alpha_3=\alpha_6=0$) the tangent vectors have the form 
\be \label{DF4} D_x F=\Phi^{-1} A\Phi, \qquad D_y F=\Phi^{-1} B \Phi\ee
with 
\bea\label{AB4} A=\alpha_4 D_x(xL), \qquad  B=\alpha_4x (D_xM)+\alpha_5 D_y(yM).\eea		
It is a straightforward computation that matrices $A$ and $B$ \eref{AB4} satisfy the condition \eref{AB} whenever the Lax equation holds \eref{cc}. From \eref{DF4}, we can integrate and find the immersion function 
\be\label{F45} F=\alpha_4x\Phi^{-1}L\Phi+\alpha_5y\Phi^{-1}M\Phi. \ee
So we have, 
\bea\fl  D_xF&=&\alpha_4\Phi^{-1}\left(L-xL^2+xD_xL+xL^2\right)\Phi+\alpha_5\Phi^{-1}\left(-LM+D_xM+ML\right)\Phi\nn
\fl &=& \alpha_4\Phi^{-1} D_x(xL)\Phi \nonumber,\eea
and 
\bea\fl  D_yF&=&\alpha_4x\left(-ML+D_yL+ML\right)\Phi+\alpha_5\Phi^{-1}\left(M-yM^2+yD_yM+yM^2\right)\Phi\nn
\fl &=& \alpha_4x\Phi^{-1} D_xM\Phi+\alpha_5\Phi^{-1}D_y(yM)\Phi,\nonumber\eea
whenever the wave function $\Phi$ satisfies the LSP \eref{lsp} and the potential matrices satisfy the Lax equation \eref{cc}. 
This shows that the immersion function $F$ can be integrated as \eref{F45} and that the vector fields 
\be \vec{v}_{Q_4}=D_x(xL)\frac{\partial}{\partial L}+x(D_xM)\frac{\partial }{\partial M},\ee
and 
\be \vec{v}_{Q_5}=M\frac{\partial}{\partial M}\ee
are generalized symmetries of the Lax equation \eref{cc}. 

Finally, we demonstrate the proof of the integrated form of the surface for the final term (when $\alpha_1=\ldots=\alpha_5=0$), as shown in a similar way in \cite{GrundPost2011a}. Suppose that $\vec{v}_Q$ is a generalized symmetry of an ODE written in Lax representation \eref{cc}. Then, the matrices 
\be\label{ABv} A=pr\vec{v}_Q(L),\qquad B=pr\vec{v}_Q(M),\ee
satisfy condition \eref{AB}
\bea\fl  D_y A-D_xB+[A,M]+[L,B]=\nn
= D_y(pr\vec{v}_Q L)-D_x(pr\vec{v}_QM)+[pr\vec{v}_Q L,M]+[L,pr\vec{v}_Q M]\nn
=pr\vec{v}_Q\left(D_yL-D_xM+[M,L]\right).\label{deteq}\eea
if and only if $\vec{v}_Q$ is a generalized symmetry of \eref{cc}. Here, we have used the fact that a generalized vector field in evolutionary form commutes with total derivatives \cite{Olver}
\be [pr\vec{v}_Q,D_x]=0, \qquad [pr\vec{v}_Q,D_y]=0,\ee
and that the potential matrices $L$ and $M$ do not explicitly depend on $y$ (ie. \eref{Dy0} holds). 
Thus, there exists a $\g$-valued immersion function $F$ for a surface with tangent vectors 
\be D_xF=\Phi^{-1} (pr\vec{v}_Q)\Phi, \qquad D_y\Phi=\Phi^{-1} (pr\vec{v}_QM)\Phi,\ee
as long as the matrices $A$ and $B$ \eref{ABv} are linearly independent. Furthermore, the immersion $F$ can be integrated explicitly as 
\be \label{Fv} F=\Phi^{-1} pr\vec{v}_Q\Phi,\ee
if and only if the vector field $\vec{v}_Q$ is a generalized symmetry of the LSP \eref{lsp}. That is, 
\bea D_x F&=&-\Phi^{-1} L (pr\vec{v}_Q\Phi)+\Phi^{-1}D_x(pr\vec{v}_Q\Phi)\nn
&=&-\Phi^{-1} L (pr\vec{v}_Q\Phi)+\Phi^{-1}pr\vec{v}_Q(D_x\Phi)\nn
&=&-\Phi^{-1}pr\vec{v}_Q(L\Phi)+\Phi^{-1}(pr\vec{v}_QL)\Phi+\Phi^{-1}pr\vec{v}_Q(D_x\Phi)\nn
&=& \Phi^{-1}(pr\vec{v}_QL)\Phi,\qquad \mbox{ whenever } pr\vec{v}_Q\left(D_x\Phi-L\Phi\right)=0,\nonumber \eea
and the proof is similar for $D_yF.$ Thus, the immersion function $F$ can be integrated as \eref{Fv} if and only if the vector field $\vec{v}_Q$ in evolutionary form is a generalized symmetry of the LSP \eref{lsp} and ODE in Lax representation \eref{cc}. 

Hence, for the system composed of equations \eref{cc} and \eref{lsp}, the corresponding formula for immersion \eref{Ffg} becomes (up to an additional $\g$-valued constant)
\be\label{Fext} F=\Phi^{-1}\left( \alpha_1 D_x+\alpha_2a(\lambda)\frac{\partial }{\partial \lambda} +\alpha_3 S+\alpha_4xL+\alpha_5 yM +pr\vec{v}_Q\right)\Phi,\ee
where $\vec{v}_Q$ is a generalized symmetry of both equations \eref{cc} and \eref{lsp}. 

In the next section, we consider second-order autonomous ODEs and their Lax pairs. The  wave function for the associated LSP is given explicitly and can be used for the purpose of constructing soliton surfaces. 

\section{Second-order autonomous equations}
Let us consider a second-order, autonomous  differential equation given by 
\be \label{uxx} u_{xx}=\frac12 f'(u), \qquad f'(u)=\frac{\partial }{\partial u}f(u) \ee
for some function $f'(u).$ Equation \eref{uxx} admits the first integral 
\be \label{ux}  u_x=\epsilon \sqrt{f(u)},\quad \epsilon^2=1, \ee
and its solutions are known  and  satisfy
\be \label{uint}\int \frac{du}{\epsilon \sqrt{f(u)}}=\epsilon(x-x_0), \qquad \ x_0 \in \mathbb{R}.\ee 
Note that, in the case when $f(u)^{-1/2}=R(u, \sqrt{P(u)})$ is a rational function of its arguments and  $P(u)$ is a polynomial of degree 3 or 4, the function $u$ which solves \eref{uxx} is the inverse of an elliptic integral \cite{ByrdFriedman}. Note also that the constant of integration, in the first integral \eref{ux}, can be absorbed since the function $f$ is an arbitrary function of $u.$

The ODE \eref{uxx} admits the following Lax pair
\[D_x \Phi=L\Phi, \qquad D_y\Phi=M\Phi,\]
where the potential matrices 
\be  \label{LM}L= \frac12\left[\ba{cc} 0 & \frac{f'(u)}{u+\lambda}-\frac{f(u)-g(\lambda)}{(u+\lambda)^2} \\ 1 & 0 \ea \right],\quad  M=\left[ \ba{cc} u_x &-\frac{f(u)-g(\lambda)}{u+\lambda}\\ u+\lambda & -u_x \ea \right]\ee
take values in the Lie algebra $sl(2, \mathbb{R})$ and are rational functions of the spectral parameter whenever $g(\lambda)$ is a rational function of $\lambda.$ Note that $det(M)=-g(\lambda)$  and the choice \[g(\lambda)=f(-\lambda)\] makes the potential matrices $L$ and $M$ polynomial in $u$ whenever $f(u)$ is a polynomial in $u$.
In what follows, we call $g(\lambda)$ the discriminant.

The goal is to construct surfaces in the Lie algebra $\g$ by the Fokas-Gel'fand procedure for the general form of the ODE \eref{uxx}. For this purpose, we solve the LSP \eref{lsp} and find explicitly the most general form of the wave function $\Phi=(\Phi_{ij})\in SL(2, \mathbb{R})$ with components \cite{GrundPost2011b}
\bea \label{phipmi}  \Phi_{11}=\frac{1}{\sqrt{2}}( \phi_{1+} +\phi_{1-}), \qquad & \Phi_{12}=\frac{-1}{2\sqrt{g}}( \phi_{1+} -\phi_{1-})\\
     \Phi_{21}=\frac{1}{\sqrt{2}}( \phi_{2+} +\phi_{2-}), \qquad & \Phi_{22}=\frac{1}{2\sqrt{g}}(\phi_{2+} -\phi_{2-}), \eea 
 where
\bea \phi_{1\pm}=\frac{\pm \sqrt{g(\lambda)}+u_x}{\sqrt{u+\lambda}}\Psi_{\pm},\\
\phi_{2\pm}={\sqrt{u+\lambda}}\Psi_{\pm},\\
 \label{phipmf}  \Psi_{\pm} =\exp \left[\pm \sqrt{g(\lambda)}\left( y+\int\frac{dx}{2(u+\lambda)}\right)\right].\eea

The generalized vector field in evolutionary form  
\be \vec{v}_Q =Q[u]\frac{\partial }{\partial u}\ee
is a generalized symmetry of the ODE \eref{uxx} if and only if 

\be \prw \left(u_{xx}-\frac12 f'(u)\right)=0, \qquad \mbox{ whenever } u_{xx}-\frac12 f'(u)=0\ee
holds.
The determining equation for $Q$ is
\be \label{detQ} D_x^2Q-\frac12f''(u)Q=0, \quad  \mbox{ whenever } u_{xx}-\frac12 f'(u)=0.\ee
Here $f'(u)$ and $f''(u)$ are the first and second derivatives of $f$ with respect to $u$. 
It is straightforward to verify that $Q=u_x$ is a solution of the determining equations \eref{detQ}.  Therefore, for an arbitrary function $f(u)$, the vector field $\vec{v}_{u_x} $ is a symmetry of \eref{cc} and \eref{lsp}, since the  the prolongation of $\vec{v}_{u_x}$ acts as a total derivative on functions which do not depend explicitly on $x$, as is the case for both \eref{cc} and \eref{lsp}. 
Consequently, we can apply the Fokas-Gel'fand procedure with ODE \eref{cc} admitting a Lax representation \eref{lsp} given by the formula \eref{Fext} with tangent vectors \eref{tanF} for matrices $A$ and $B$ as in \eref{Aext} and \eref{Bext}.

Two possible choices for a scalar product can be introduced on the tangents to the surface $F\in \glr$ with the basis given by 
\be \label{basis} e_1=\left[\ba{cc}1&0\\ 0&-1 \ea \right], \quad e_2=\left[\ba{cc}0&1\\ 1&0 \ea \right], \quad e_3=\left[\ba{cc}0&-1\\ 1&0 \ea \right].\ee
In the first case, we decompose the matrix into the $sl(2,\mathbb{R})$ basis and then use the standard Euclidean metric. The inner product and its norm in Euclidean space are given by 
\be \label{Euc} \langle X, Y \rangle=X^iY^i, \qquad || X||=\sqrt{X^iX^i},\ee
where $X=X^ie_i$, $Y=Y^je_j \in sl(2,\mathbb{R})$ $i,j=1,2,3$. With the inner product \eref{Euc} the surfaces are Riemannian manifolds. 

In the second case, we use the Killing form on $\glr,$ which is given by \cite{DoCarmo, Helgason}
\be B(X,Y)=\frac12 \tr(XY).\ee
In terms of the basis \eref{basis}, the matrices $X,Y\in \glr $ and the Killing form can be represented as follows
\bea B(X^i, Y^j)=X^iB_{ij}Y^j, \\
 B_{ij}=\left[ \ba{ccc} 1 & 0 & 0\\ 0 & 1 &0\\ 0& 0&-1\ea \right] .\eea
 So, the Killing form has signature $(2,1)$ and induces a pseudo-Euclidean metric. The surfaces $F\in \glr$ are pseudo-Riemannian manifolds. 
 
Let us now explore certain geometric characteristics of the surfaces immersed in the $\glr$ algebra. These geometric properties include the fundamental forms, mean and Gaussian curvatures. For example,  the first fundamental form for the surfaces $F^{ST}$ and $F^{u_x}$ for any function $f(u)$  with the pseudo-Euclidean metric are given by  
\be I_B(F^{ST})=\left(2\frac{f-g}{(u+\lambda)^3}-\frac{f'-g'}{(u+\lambda)^2}\right)dxdy+2\left(\frac{g'}{v+\lambda}+\frac{f-g}{(u+\lambda)^2}\right)dy^2
,\ee
and 
\be\fl  I_B(F^{u_x})=\left(\frac{ff''}{u+\lambda}-\frac{2ff'}{(u+\lambda)^2}+\frac{2f(f-g)}{(u+\lambda)^3}\right)dxdy+\left(\frac{f'^2}{2}-\frac{2ff'}{u+\lambda}+\frac{2f(f-g)}{(u+\lambda)^2}\right)dy^2.\ee
Note that in both cases, the first fundamental forms admit null vectors in the $dx$ direction. 

\section{The static $\phi^4$ field equation and its soliton surfaces}
We now present an example which illustrates the theoretical considerations. We intend to discuss in detail the construction of static and translation-invariant solutions of the $\phi^4$-field equations \cite{11}
\be \label{phi4} \Delta M=\frac{A}{2D} M+\frac{B}{2D} M^3, \qquad 0< A, B, D, \in \mathbb{R}, \ee
where $\Delta$ denotes the Laplace operator on variables $(x,y,z)$. 
For the purpose of this investigation, we limit ourselves to the three-dimensional Lie subalgebra spanned by $\{ L_1, P_2, P_3\}$, where the infinitesimal generators of rotation $L_1$ and translations $P_2$ and $P_3$ are 
\be L_1=y\frac{\partial}{\partial z}-z\frac{\partial}{\partial y}, \qquad P_2=\frac{\partial}{\partial y}, \qquad P_3=\frac{\partial }{\partial z}. \ee
The invariant solutions are of the form \cite{12}
\be M(x)=u(\xi), \qquad \xi=\overline{e}(\vec{x}-\vec{x}_0), \qquad |\vec{e}|^2=1, \ee
where $\vec{x}_0$ and $\vec{e}$ are constant vectors and the variable $\xi$ is obtained by applying the rotation $L_1$ and translations $P_2, P_3$ to the symmetry variable $\xi=x$. The translationally symmetric solution $u(\xi)$ satisfies the second-order equation 
\be\label{xn} u_{xx} =-2k_2u^3+(k_2-k_1)u,\ee
where 
\be k_1=\frac{-1}{2D}(A+\frac{B}{2}), \qquad k_2=\frac{-B}{4D}.\ee
Integrating once, \eref{xn} the first integral is 
\be \label{xnx} (u_x)^2=(1-u^2)(k_1+k_2u^2).\ee
For specific choices of constants $k_1$ and $k_2$, we obtain different Jacobian elliptic functions $sn$ $cn $ and $dn$ \cite{BriBoubook, ByrdFriedman}.
\be \ba{ccc}
k_1 & k_2& \mbox{Solutions of \eref{xnx}}\\
\mr
1 & -k^2 & sn(x, k) \\ 
k'^2 &  k^2 & cn(x,k) \\
-k'^2 &  1 &  dn(x,k) \ea \ee
The moduli $k$ of the elliptic functions are chosen in such a way that $k'^2+k^2=1$ and $0\leq k, k' \leq 1.$ This ensures that the elliptic solutions possess one real and one purely imaginary period. 

The potential matrices $L$ and $M$ take values in the $\glr$ Lie algebra and are polynomial in $u$ of third degree \cite{27}
\bea\label{Mjac} M=\left[\ba{cc} u_x& (u-\lambda)(k_2(u^2+\lambda^2)+k_1-k_2)\\ u+\lambda& -u_x\ea \right]\in sl(2,\mathbb{R}),\\
L=\frac12\left[\ba{cc} 0&  -3k_2u^2+2\lambda k_2u +k_1-k_2-k_2\lambda^2 \\1 &0\ea\right]\in sl(2,\mathbb{R}),\label{Ljac}\eea 
where the discriminant is chosen as 
\be g(\lambda)=f(-\lambda)=(1-\lambda^2)(k_1+k_2\lambda^2).\ee 
Solving the LSP \eref{lsp}  with potential matrices given by \eref{Mjac} and \eref{Ljac}, the most general form of the wave function is \cite{GrundPost2011b}
\begin{equation}\label{Phif} \fl \Phi=\left[\ba{cc} \frac{(\sqrt{g(\lambda)}-u_x)\Psi_+-(\sqrt{g(\lambda)}+u_x)\Psi_-}{2\sqrt{u+\lambda}},& \frac{(\sqrt{g(\lambda)}+u_x)\Psi_--(\sqrt{g(\lambda)}-u_x)\Psi_+}{2\sqrt{g(\lambda)}\sqrt{u+\lambda}}\\ \frac{\sqrt{u+\lambda}(\Psi_++\Psi_-)}{2}, & \frac{\sqrt{u+\lambda}(\Psi_--\Psi_+)}{2\sqrt{g(\lambda)}} \ea\right]\in SL(2,\mathbb{R}),\end{equation}
where  
\bea \fl \Psi_\pm &=\exp\left[\pm\sqrt{g(\lambda)}\left(y+ \frac{\epsilon}{\lambda\sqrt{k_1}}\Pi\left(u,\frac{1}{\lambda^2}, \sqrt{\frac{-k_2}{k_1}}\right)+c_0\right)\right]\nn
\fl &\times\left[\frac{2\sqrt{g(\lambda)}\sqrt{(1-u^2)(k_1+k_2u^2)}+(k_2-k_2-2k_2\lambda^2)u^2+(k_2-k_1)\lambda^2+2k_1}{2\sqrt{g(\lambda)}\sqrt{(1-u^2)(k_1+k_2u^2)}-(k_2-k_2-2k_2\lambda^2)u^2-(k_2-k_1)\lambda^2-2k_1}\right]^{\mp \frac{\epsilon}4}.\eea
where $c_0$ is a real integration constant and $\Pi\left(u,a, b\right)$ is the normal elliptic integral of the third kind, see e.g. \cite{ByrdFriedman} 
\be\label{Pi} \Pi\left(u,\alpha^2,k \right)=\int_0^x\frac{dt}{(1-\alpha^2t^2)\sqrt{1-t^2}\sqrt{1-k^2t^2}}.\ee In the graphs below, we choose the integration constant $c_0$ so that $\Psi_\pm(0,0)=1.$

The surfaces associated with the elliptic equations \eref{xn} are given by the formula \eref{Fext}. As an example, we consider separately three cases of the Jacobian elliptic function $sn(x,k)$. When the discriminant $g(\lambda)>0$, the surfaces display exponential type behavior and when $g(\lambda)<0$ the surfaces behave like trigonometric functions (see figure 1).

In the pseudo-Euclidean metric defined by the Killing form, the first and second fundamental forms for the surfaces $F^{ST}$ and $F^{Q}$ with $Q=u_x$ are 
\bea \fl I(F^{ST})=-k^2(u-\lambda)dxdy+\left(2k^2u^2-4\lambda k^2u+6k^2\lambda^2-2k^2-2\right)dy^2\nn
\fl II(F^{ST})=2^{-1/2}k^2(u-\lambda)dx^2-2^{1/2}k^2(u^2-\lambda^2)dxdy\nn
+2^{3/2}\left({k}^{2}{u}^{3}-\lambda\,{k}^{2}{u}^{2}+ \left( {k}^{2}{\lambda}^{2}-{k}^{2}-1 \right) u+{k}^{2}{\lambda}^{3}\right)dy^2\nn
 \fl I(F^{Q})=k^2(u^2-1)(k^2u^2-1)(3u-\lambda)dxdy\nn
+2\Bigg(\,{k}^{4}{u}^{6}+4\,{k}^{4}{u}^{5}\lambda-2\,{k}^{4}{u}^{4}{\lambda}^
{2}-4\,{k}^{2}\lambda\, \left( 1+{k}^{2} \right) {u}^{3}\nn \qquad 
+2\,{k}^{2}
 \left( -3+{\lambda}^{2}+{k}^{2}{\lambda}^{2} \right) {u}^{2}+4\,
\lambda\,{k}^{2}u+2-2\,{k}^{2}{\lambda}^{2}+2\,{k}^{2}\Bigg)dy^2\nn
\fl II(F^{Q})=\frac{u_xk^2(\lambda-3u)}{\sqrt{2}}dx^2+\sqrt{2}u_xk^2(\lambda-3u)(u+\lambda)dxdy\nn
 +\frac {2\sqrt {2} \left( u+\lambda \right)}{u_x} \Bigg( -1+{k}^{2}{\lambda}^{2}+3\,{k}^{2}{u}^{2}-
{k}^{2}-2\,\lambda\,{k}^{2}u-{u}^{2}{k}^{2}{\lambda}^{2}-2\,{k}^{4}{u}
^{5}\lambda\nn\qquad +2\,{u}^{3}\lambda\,{k}^{2}-{k}^{4}{u}^{2}{\lambda}^{2}+2\,
{k}^{4}\lambda\,{u}^{3}+{k}^{4}{u}^{4}{\lambda}^{2}-{k}^{4}{u}^{6} \Bigg)dy^2, \nonumber\eea
The normals are 
\bea  N(F^{ST})=\frac{1}{\sqrt{2}}e_1\nn
         N(F^{Q})=\left[ \begin {array}{cc} -1/2\,\sqrt {2}&{\frac { \left( 2\,{k}^{2}{
u}^{2}-{k}^{2}-1 \right) u\sqrt {2}}{u_{{x}}}}\\\noalign{\medskip}0&1/
2\,\sqrt {2}\end {array} \right] \nonumber\eea
and the Gaussian and mean curvatures are 
\bea \fl  K(F^{ST})=2\, \left( 2\,{k}^{2}{u}^{2}-{k}^{2}-1 \right)  \left( \lambda-u
 \right) {k}^{2}u\nn
\fl H(F^{ST})={k}^{2}{\lambda}^{2}-2\,\lambda\,{k}^{2}u+3\,{k}^{2}{u}^{2}-{k}^{2}-1\nn
\fl K(F^{Q})=2\,{k}^{2} \left( u+\lambda \right)  \left( \lambda-3\,u \right) 
 \left( -1-{k}^{2}-3\,{k}^{2}{u}^{4}+6\,{k}^{2}{u}^{2}+2\,{k}^{4}{u}^{
6}-3\,{k}^{4}{u}^{4} \right) \nn
\fl H(F^{Q})=\sqrt {2}\left( \lambda-3\,u
 \right) u_{{x}}{k}^{2}\Bigg(-5\,{k}^{4}{u}^{6}-2\,{k}^{4}{u}^{5}\lambda+{k}^{2} \left( 6+{k}^{2}{
\lambda}^{2}+6\,{k}^{2} \right) {u}^{4}\nn+2\,{k}^{2}\lambda\, \left( 1+{
k}^{2} \right) {u}^{3}-{k}^{2} \left( 9+{k}^{2}{\lambda}^{2}+{\lambda}
^{2} \right) {u}^{2}-2\,\lambda\,{k}^{2}u+{k}^{2}{\lambda}^{2}+1+{k}^{
2} \Bigg)  \nonumber. \eea 
Note that the second fundamental form and the Gaussian curvature for the surface $F^{ST}$ in the pseudo-Euclidean metric coincide with those given for the surface in the Euclidean metric \cite{GrundPost2011b}. Also, the surface $F^{ST}$ lies in a plane in the moving frame defined by conjugation with respect to the wave function $\Phi.$ Graphs of these surfaces can be found in \cite{GrundPost2011b}. 

Finally, we consider the surfaces associated with dilation symmetry (i.e. with constant $\alpha_4$ in \eref{Fext}). The surface is given by 
\be F^{4}=\Phi^{-1}xL\Phi,\ee
and the first fundamental form is 
\bea \fl I(F^{4})=\left(3/2\,{k}^{2}{u}^{2}-{k}^{2} \left( \lambda-3\,xu_{{x}} \right) u+1/2\,
{k}^{2}{\lambda}^{2}-u_{{x}}x{k}^{2}\lambda-1/2\,{k}^{2}-1/2\right)dx^2\nn
+k^2x^2(u^1-2)(k^2u^2-1)(3u-\lambda)dxdy\nn
+2\,{x}^{2} \Bigg({k}^{4}{u}^{6}+2\,{k}^{4}{u}^{5}\lambda-{k}^{4}{u}^{4}{\lambda}^{2}-2
\,{k}^{2}\lambda\, \left( 1+{k}^{2} \right) {u}^{3}\nn \qquad \qquad 
+{k}^{2} \left( -3+
{\lambda}^{2}+{k}^{2}{\lambda}^{2} \right) {u}^{2}+2\,\lambda\,{k}^{2}
u+1-{k}^{2}{\lambda}^{2}+{k}^{2}\Bigg)dy^2.\nonumber\eea
 The other geometric characteristics are directly computable but are too involved to write out in an illustrative fashion. Below we gives graphs of these surfaces in figures 1. 

\begin{figure}\label{gneg}
\begin{center}$
\begin{array}{cc}

\includegraphics[width=2.5in]{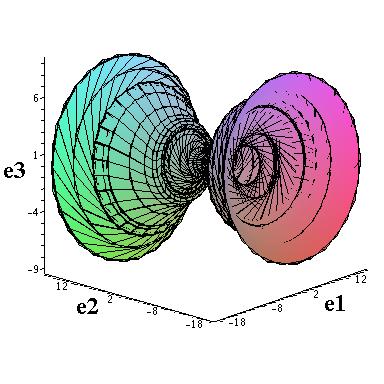} & \includegraphics[width=2.5in]{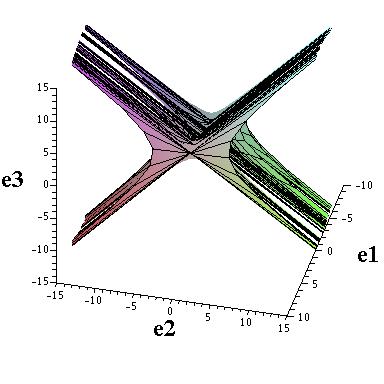}\\
F^{4}: \ \lambda=1.2, \ k=0.5, \ g(\lambda)<0 &\lambda=0.5, \ k=0.5, \ g(\lambda)>0\end{array}$
\caption{Surfaces $F^{4}$ for $u=sn(x,k)$  and  $x$ and $y \in [-8,8]$. The axes indicate the components of the immersion function in the basis \eref{basis}  }
\end{center}
\end{figure}

\section{Final remarks and future perspectives}
The main goal of this paper is to discuss an adaptation of the Fokas-Gel'fand procedure for constructing exact, analytic soliton surfaces associated with integrable ODEs admitting a Lax representation, as presented in \cite{GrundPost2011b}. In our investigation of the immersion formula for 2D-surfaces immersed in a Lie algebra we have proceeded in the following manner:
\begin{itemize}
\item[1] We have shown, as in the PDE case \cite{GrundPost2011a}, the problem of immersion requires the examination of conformal symmetries in the spectral parameter, gauge symmetries of the LSP and generalized symmetries of the associated ODE model and its LSP (see the formula for immersion \eref{Ffg}). We have also demonstrated addition terms in the immersion formula, associated with translation in the independent variable $x$ and dilation symmetry (see the extended formula for immersion \eref{Fext}). 
\item[2] We have constructed a Lax pair for a second-order ODE depending on an arbitrary function $f(u)$ which includes, among others, the case of elliptic equations. 
\item[3] We found explicitly the most general form of the wave function.
\item[4] From the general solution for the wave function, we have constructed 2D-surfaces in the Lie algebra $\glr$ by analytic methods for the general form of the ODE $(u_x)^2=f(u),$ with arbitrary $f(u)$. 
\item[5] Next, we identified a generalized symmetry of the considered ODE and showed that the immersion function can be explicitly integrated as in \eref{Fext}. 
\item[6] The procedure was illustrated for translationally invariant solutions of the Landau-Ginzburg equation leading to the Jacobian elliptic equation. We have given some geometric characterizations such as the first and second fundamental forms of the surfaces as well as the mean and Gaussian curvatures. Additionally, we have provided graphs of the surfaces for different range of parameters leading to diverse types of surfaces. 
\end{itemize}

In the next stage of this research, using a group theoretical technique, the authors plan to consider soliton surfaces immersed in Lie algebras associated with the KdV and MKdV equations 
\be u_t+u_{xxx}+6u_xu=0, \qquad u_t+u_{xxx}-6u_xu^2=0\ee
respectively. The stationary states of these equations, when $u_t=0$, are given by Jacobian elliptic functions. Thus, a natural extension would be to compare the surfaces associated with PDEs as presented in \cite{GrundPost2011a} to the ODE surfaces associated with their stationary states (here the auxiliary variable $y$ becomes $t$). For this purpose, we can use the exact solutions of the wave function \eref{phipmi} for stationary states to expand analytically for solutions in the neighborhood of stationary states and look for solutions of the wave function in the PDE case. 

Another avenue for investigation is the possibility of performing asymptotic analysis for the study of PDE surfaces using the ODE surfaces as approximations. A further investigation might be to consider how to apply the recurrence operator to generalized symmetries of the KdV and MKdV model in order to obtain recurrence relations for the surfaces. These and other issues will be addressed in our future work.

\ack  The research reported in this paper is supported by NSERC of Canada. S Post acknowledges a postdoctoral fellowship provided by the Laboratory of Mathematical Physics of the Centre de Recherches Math\'ematiques, Universit\'e de Montr\'eal. 
\appendix

\end{document}